\documentclass[12pt]{spieman}  
\usepackage{amsmath,amsfonts,amssymb}
\usepackage{graphicx}
\usepackage{setspace}
\usepackage{tocloft}
	
\usepackage{xcolor}
\usepackage{soul}

\renewcommand{\vec}[1]{\mathbf{#1}}
\usepackage{float}

\usepackage{hyperref}
\usepackage{adjustbox}
\usepackage{multirow}
\usepackage{multicol}
\usepackage{makecell}
\usepackage{subcaption}
\usepackage{lineno}

\title{Learning stochastic object models from medical imaging measurements by use of advanced ambient generative adversarial networks}

\author[a]{Weimin Zhou}
\author[b]{Sayantan Bhadra}
\author[c]{Frank J. Brooks}
\author[c,d,e]{Hua Li}
\author[c,*]{Mark A. Anastasio}
\affil[a]{University of California Santa Barbara, Department of Psychological and Brain Sciences, Santa Barbara, CA, USA}
\affil[b]{Washington University in St. Louis, Department of Computer Science and Engineering, St. Louis, MO, USA}
\affil[c]{University of Illinois at Urbana-Champaign, Department of Bioengineering, Urbana, IL, USA}
\affil[d]{Washington University in St. Louis, Department of Radiation Oncology, Saint Louis, MO, USA}
\affil[e]{Cancer Center at Illinois, Urbana, IL, USA}

\cftpagenumbersoff{figure}
\cftpagenumbersoff{table} 
\begin{document} 
\maketitle

\vspace{-0.8cm}
\begin{abstract}

\noindent \textbf{Purpose:} In order to objectively assess new medical imaging technologies via computer-simulations, it is important to account for the variability in the ensemble of objects to-be-imaged.
This source of variability can be described by stochastic object models (SOMs). 
It is generally desirable to establish SOMs from experimental imaging measurements acquired by use of a well-characterized imaging system, but this task has remained challenging. 

\noindent \textbf{Approach:} 
A generative adversarial network (GAN)-based method that employs AmbientGANs with modern progressive or multi-resolution training approaches is proposed.
 AmbientGANs established by use of the proposed training procedure are systematically validated in a controlled way by use of computer-simulated magnetic resonance imaging (MRI) data corresponding to a stylized imaging system.
Emulated single-coil experimental MRI data are also employed to demonstrate the methods under less stylized conditions.

\noindent \textbf{Results:} 
The proposed AmbientGAN method can generate clean images when the imaging measurements are contaminated by measurement noise.
When the imaging measurement data are incomplete, the proposed AmbientGAN can reliably learn the distribution of the measurement components of the objects.

\noindent \textbf{Conclusions:}  
Both visual examinations and quantitative analyses, including task-specific validations by use of the Hotelling observer, demonstrated that the proposed AmbientGAN method holds promise to establish realistic SOMs from imaging measurements.

\end{abstract}

\vspace{-0.1cm}

\vspace{-0.1cm}
{\noindent \footnotesize\textbf{*}Mark A. Anastasio,  \linkable{maa@illinois.edu} }

\begin{spacing}{1}   
\vspace{-0.1cm}
\section{Introduction}
\vspace{-0.1cm}
\label{sect:intro}  
Computer-simulation remains an important approach for the design and optimization
of imaging systems. Such approaches can permit the exploration, refinement, and assessment
 of a variety of system designs that would be infeasible through experimental studies alone \cite{abadi2020virtual,badano2017silico,wilson2000reconstruction}.
In the field of medical imaging, it has been advocated that imaging systems
 and reconstruction algorithms
 should be assessed and optimized
by use of objective measures of image quality (IQ) that quantify the performance of an observer at specific diagnostic tasks \cite{myers1993rayleigh,wagner1985unified,barrett1993model,barrett2013foundations,anastasio2010analysis}.
To accomplish this,  all sources of variability in the measured data should  be
accounted for.
One important source of variability that can significantly limit observer performance
is variation in the objects to-be-imaged \cite{rolland1992effect}.
This source of variability can be described by stochastic object models (SOMs)
 \cite{kupinski2003experimental}.
A SOM is a  generative model that can be employed to produce an ensemble of to-be-imaged objects 
that possess prescribed statistical properties.

Available SOMs include  texture models of mammographic images with
clustered lumpy backgrounds \cite{bochud1999statistical}, simple lumpy background models
\cite{rolland1992effect}, and more realistic 
anatomical phantoms that can be randomly perturbed \cite{segars2008realistic}.
A variety of other computational phantoms~\cite{segars2002study,xu2014exponential,zankl2010gsf,collins1998design,caon2004voxel, zu2005vip, segars2008realistic,li2009methodology},
either voxelized or mathematical, have been proposed for medical imaging simulation,
 aiming to provide a practical solution to characterize object variability.
However, the majority of these were established by use of image data corresponding to only a few subjects.
Therefore, they may not accurately describe the statistical properties
of the  ensemble of objects that is relevant to an imaging system optimization task. 
 A variety of anatomical shape models have also been proposed to describe both the common geometric features
and the geometric variability among instances of the population for shape
 analysis applications~\cite{cootes1995active,cootes2015active,heimann2009statistical,ferrari2010images,shen2012detecting,gordillo2013state,
tomoshige2014conditional,ambellan2019automated}.
To date, these have not been systematically explored for the purpose of
 constructing SOMs that capture realistic anatomical variations
for use in imaging system optimization.

In order to establish SOMs that capture realistic textures and anatomical variations,
it is desirable to utilize experimental imaging data.
By definition, however, SOMs should be independent of the imaging system, measurement noise
 and any reconstruction method employed. In other words, they should provide an \emph{in silico}
representation of the ensemble
of objects to-be-imaged and not estimates of them that would be indirectly measured
or computed by imaging systems.  To address this need, 
Kupinski \emph{et al.} \cite{kupinski2003experimental}
proposed an explicit generative model for describing object statistics
 that was trained by use of noisy imaging measurements and a computational model of
a well-characterized imaging system~\cite{kupinski2003experimental}.
However,
applications of this method have been limited to situations
 where the characteristic functional of the random object can be analytically determined \cite{clarkson2016characteristic},
such as with lumpy and clustered lumpy object models\cite{kupinski2005small,bochud1999statistical}.
As such, there remains an important need
to generalize the method.

Deep generative neural networks, such as generative adversarial networks
 (GANs) \cite{goodfellow2014generative}, hold great potential for establishing SOMs
that describe   finite-dimensional approximations of objects.
However, conventional GANs are typically trained by use of reconstructed
 images that are influenced by the effects of measurement noise
 and the reconstruction process.
 To circumvent this, an AmbientGAN has been proposed \cite{bora2018ambientgan} that augments
 a GAN with a measurement operator. This permits
 a generative model that describes object randomness 
 to be learned from indirect and noisy measurements of the objects
themselves.
In a preliminary study, the AmbientGAN was explored
 for establishing SOMs from imaging measurements for use in optimizing imaging systems \cite{zhou2019learning}.
 However, similar to conventional GANs, the process of training AmbientGANs
is inherently unstable.
Moreover, the original AmbientGAN cannot immediately benefit from
 robust GAN training procedures, such as progressive growing\cite{karras2018progressive},
 which limits its ability to synthesize high-dimensional images that depict accurate approximations of
objects that are relevant to medical imaging studies.

In this work, 
 modern multi-resolution training approaches, such as employed in the Progressive Growing of GANs (ProGANs) \cite{karras2018progressive} and Style-based GANs (StyGANs) \cite{karras2019style, karras2019analyzing}, are modified for use in establishing AmbientGANs with high-dimensional medical imaging measurements.  The resulting  models will be referred to 
 as  Progressive Growing AmbientGANs (ProAmGANs) and Style-AmbientGANs (StyAmGANs).
Numerical studies  corresponding to
a stylized imaging system are 
conducted to systematically investigate the proposed advanced AmbientGAN methods for establishing SOMs.
The  effects  of  noise  levels  and  the imaging operator  null  space  characteristics on model performance are assessed by use of both standard and objective measures. Emulated single-coil experimental magnetic resonance imaging data are also employed to demonstrate the method under less stylized conditions.

The remainder of this paper is organized as follows. 
In Sec.~\ref{sec:bkgd}, previous works on learning SOMs
 by employing characteristic functions and AmbientGANs 
are summarized.  
The progressive growing training strategy for GANs is also reviewed.
The proposed ProAmGAN and StyAmGAN for learning SOMs from noisy
 imaging measurements are described in Sec.~\ref{sec:ProAmGAN}.
Sections \ref{sec:num} and \ref{sec:result} describe the numerical studies and results
that demonstrate the ability of the advanced AmbientGANs to learn SOMs from
noisy imaging measurements.
Finally,  a discussion and summary of the work is presented in
Sec.~\ref{sec:conclusion}.

\section{Background}
\label{sec:bkgd}
Object properties that are imaged by medical imaging systems are inherently described by continuous functions.
However, it is common practice when performing computer-simulation studies of imaging systems to approximate the object by use of a finite-dimensional representation \cite{lewitt1990multidimensional,panin2006fully}. 
In such cases,
a discrete-to-discrete (D-D) description of 
a linear imaging system can be described as~\cite{barrett2013foundations}:
\begin{equation}\label{eq:imaging}
    \mathbf{g} = \mathbf{H}\mathbf{f} + \mathbf{n},
\end{equation}

\noindent where $\mathbf{g} \in \mathbb{R}^M$ is a vector that describes the measured image data,
$\mathbf{f} \in \mathbb{R}^N$ denotes the finite-dimensional representation of the object being imaged, 
$\mathbf{H}\in \mathbb{R}^{M\times N}$ denotes a D-D imaging operator $\mathbb{R}^N\rightarrow \mathbb{R}^M$ that maps an object in the Hilbert space $\mathbb{U}$ to the measured discrete data in the Hilbert space $\mathbb{V}$, 
and the random vector $\mathbf{n} \in \mathbb{R}^M$ denotes the measurement noise. 
Below, the imaging process described in Eq.~(\ref{eq:imaging}) is denoted as: $\mathbf{g} = \mathcal{H}_{\mathbf{n}}(\mathbf{f})$.  
In this work, it will be assumed that the D-D imaging model is a sufficiently accurate representation
of the true continuous-to-discrete (C-D) imaging model that describes a digital imaging system and the impact of model error will be neglected. Accordingly, as described below, the objective of this work will be to establish SOMs that describe the finite-dimensional vector $\mathbf{f}$.

When optimizing imaging system performance by use of objective measures of IQ, all sources
of randomness in $\mathbf{g}$ should be considered. In diagnostic imaging applications,
object variability is an important factor that limits observer performance.
In such applications, the object  $\mathbf{f}$ should be described as a random vector
that is characterized by a multivariate probability density function (PDF)
 $p(\mathbf{f})$ that specifies the
statistical properties of the ensemble of objects to-be-imaged.

Direct estimation of $p(\mathbf{f})$ is rarely tractable in medical imaging applications due 
to the high dimensionality of $\mathbf{f}$.
To circumvent this difficulty, a parameterized generative model, referred to throughout this work as
a SOM,  can be introduced and established
by use of an ensemble of experimental measurements. The generative model can be explicit
or implicit.
Explicit generative models seek to approximate $p(\mathbf{f})$, or equivalently, its
characteristic function,  from which samples
$\mathbf{f}$ can subsequently be drawn. 
On the other hand, implicit generative models do not seek to estimate $p(\mathbf{f})$ directly,
 but rather define a stochastic process that can draw samples
from $p(\mathbf{f})$ without having to explicitly specify it.  Variational autoencoders and GANs are examples
 of  explicit and implicit generative models, respectively, that have
been actively explored \cite{goodfellow2016deeplearning}.
Two previous works that sought to learn SOMs from noisy and indirect imaging measurements  by use of explicit and implicit generative
models are presented below.

\subsection{Establishing SOMs by use of explicit generative modeling: Propagation
of characteristic functionals}
\label{ssec:CFforSOM}

The first method to learn SOMs from imaging measurements was introduced by
Kupinski \textit{et al.}~\cite{kupinski2003experimental}.
In that seminal work, a C-D imaging model was considered in which a function that describes the object is mapped to a finite-dimensional image vector
$\mathbf{g}$.
For C-D operators, it has been demonstrated that 
 the characteristic functional (CFl) describing the object 
 can be readily related to the characteristic function (CF) 
of the measured data vector $\mathbf{g}$ \cite{clarkson2002transformation}.
 This provides  a
relationship between the PDFs of the object and measured image data.
In their method, an object that was parameterized by
the vector $\mathbf{\Theta}$ was considered and
analytic expressions for the CFl were utilized.
Subsequently, by use of the known imaging operator and noise model,
  the corresponding CF  was computed. The vector $\mathbf{\Theta}$
was estimated by 
 minimizing the discrepancy between this model-based CF and
an empirical estimate of the CF computed from an ensemble of noisy
imaging measurements. From the estimated CFl,
an ensemble of objects could be generated.
This method was applied to establish SOMs where the CFl of the object can
be analytically determined. 
Such cases include the
lumpy object model\cite{kupinski2005small} and
 clustered lumpy object model \cite{bochud1999statistical}.
The applicability of the method to more complicated object
models  remains unexplored.

\subsection{Establishing finite-dimensional SOMs by use of implicit generative
modeling: GANs and AmbientGANs}
\label{ssec:AmGANforSOM}

Generative adversarial networks (GANs)~\cite{goodfellow2014generative, arjovsky2017towards,arora2017do,denton2015deep,radford2015unsupervised,
salimans2016improved, li2019misGAN, shrivastava2015learning, arjovsky2017wasserstein, gulrajani2017improved, brock2018large} 
 are implicit generative models that have been actively explored to
learn the statistical properties of ensembles of images (i.e., finite-dimensional approximations
of object properties) and generate new images
that are consistent with them.
A traditional GAN consists of two deep neural networks - a  generator and a discriminator. 
The generator is jointly trained with the discriminator through an adversarial process. 
During its training process, the generator is trained to map random low-dimensional latent vectors
to higher dimensional images that represent samples from the distribution of training images.
The discriminator is trained to distinguish the generated, or synthesized, images from
 the actual training images. These are often referred to as the ``fake" and ``real" images in the
GAN literature. 
Subsequent to training, the discriminator is discarded and the generator and associated
latent vector probability distribution form as an implicit
generative model that can sample from the data distribution to produce new images.
However, images produced by  imaging systems are contaminated by measurement
noise and potentially an image reconstruction process.  Therefore, GANs trained
directly on images do not generally represent SOMs because they do not
characterize object variability alone.

\begin{figure}[H]
\centering
\includegraphics[width=0.7\linewidth]{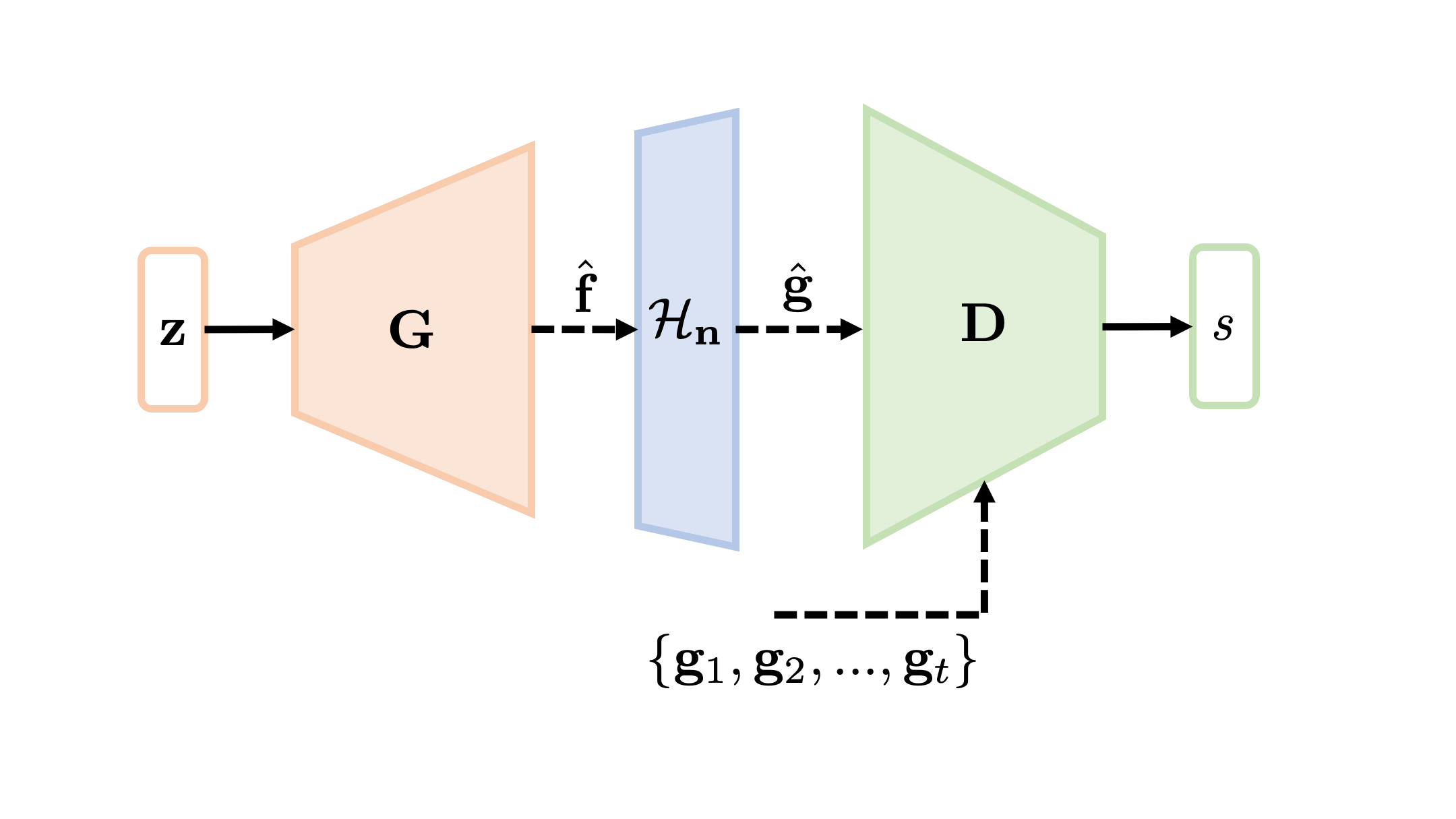}
\vspace{-0.8cm}
\caption{An illustration of the AmbientGAN architecture. The generator $\mathbf{G}$ is trained to generate objects, which are subsequently employed to simulate measurement data. The discriminator $\mathbf{D}$ is trained to distinguish ``real" measurement data from the ``fake" measurement data that are simulated by use of the generated objects.}
\label{fig:arc_AGAN}
\end{figure}

An augmented GAN architecture named AmbientGAN has been proposed~\cite{bora2018ambientgan}
that enables learning a SOM that describes the statistical properties of finite-dimensional approximations of objects
from noisy indirect measurements of the objects.
As shown in Fig.~\ref{fig:arc_AGAN}, the AmbientGAN architecture incorporates
 the measurement operator  $\mathcal{H}_{\vec{n}}$, defined in Eq.\ (\ref{eq:imaging}),
 into the traditional GAN framework.
During the AmbientGAN training process,
the generator is trained to map a random vector $\mathbf{z} \in \mathbb{R}^{k}$ described by a latent probability distribution to a generated object
$\hat{\mathbf{f}} = G(\mathbf{z}; \mathbf{\Theta}_{G})$, 
where $G: \mathbb{R}^{k} \rightarrow \mathbb{R}^N$ represents the generator network
 that is parameterized by a vector of trainable parameters $\mathbf{\Theta}_G$.
Subsequently, the corresponding simulated imaging measurements are computed as
 $\hat{\vec{g}} = \mathcal{H}_\vec{n}(\hat{\vec{f}})$.
The discriminator neural network $D: \mathbb{R}^{M} \rightarrow \mathbb{R}$,
which is parameterized  by the vector $\mathbf{\Theta}_D$,
is trained to distinguish the real and simulated imaging measurements by mapping them to a real-valued scalar $s$.
The adversarial training process 
 can be represented by the following two-player minimax game\cite{goodfellow2014generative}:
\begin{equation} \label{eq:AGAN}
\min_{\mathbf{\Theta}_G} \max_{\mathbf{\Theta}_D} V(D,G) =  {E_{\vec{g}\sim  p(\vec{g}) }} [l\left(D(\vec{g}; \mathbf{\Theta}_D)\right)]
 + {E_{\hat{\vec{g}} \sim  p(\hat{\vec{g}}) }} [l(1- D\left( \hat{\vec{g}}; \mathbf{\Theta}_D \right) )],
\end{equation}
where $l(\cdot)$ represents a loss function.  
When the distribution of objects $p(\vec{f})$ uniquely
 induces the distribution of imaging measurements $p(\vec{g})$, i.e., when the 
imaging operator is injective, and the minimax game achieves the global optimum, 
the trained generator can be employed to produce object samples drawn from $p(\vec{f})$~\cite{goodfellow2014generative, bora2018ambientgan}.   
\vspace{-0.5cm}

Zhou \textit{et al.} have demonstrated the ability of the AmbientGAN to learn a
simple SOM corresponding to a lumpy object model that could be 
employed to produce small ($64\times 64$) object samples \cite{zhou2019learning}.
However,
adversarial training is known to be unstable
and the use of AmbientGANs to establish realistic and large-scale SOMs has, to-date, been limited.

\subsection{Advanced GAN training strategies}
\label{sec:ProGAN}

A novel training strategy for GANs---progressive growing of GANs (ProGANs)---has been recently developed to improve the stability of the GAN training process~\cite{karras2018progressive} 
and hence the ability to learn  generators that sample
from distributions of high-resolution images.
GANs are conventionally trained directly on full size images through the entire training process.
 In contrast, ProGANs adopt a multi-resolution approach to training.
Initially, a generator and discriminator are trained by use of down-sampled (low resolution) training
images.
During each subsequent training stage, higher resolution versions of the original training images are employed
to train progressively deeper discriminators and generators, continuing until a final version of the
  generator
is trained by use of the original high-resolution images. A similar progressive training procedure
is employed in the StyleGAN framework \cite{karras2019style}.
More recently, an advanced GAN training strategy---StyleGAN2---has been developed to further improve the image quality of the synthesized images\cite{karras2019analyzing}. Although, StyleGAN2 does not employ the progressive growing strategy, the generator does 
 make use of multiple scales of image generation via skip connections between lower resolution generated images to the final generated image\cite{karras2019analyzing}.
While these advanced training strategies have found widespread success on training GANs, 
they cannot be directly used to train AmbientGANs for establishing SOMs from medical imaging measurements.
This is because these GAN training procedures and architectures are designed to train the generator that produces images in the same Hilbert space as the training images.
However, medical imaging measurements $\vec{g}$ that are used as training data of AmbientGANs 
are typically indirect representations of to-be-imaged objects $\vec{f}$ and generally reside in a different Hilbert space than the generator-produced objects $\hat{\vec{f}}$.
For example,
in magnetic resonance imaging (MRI),
the to-be-imaged objects reside in a real Hilbert space while the k-space measurements reside in a complex Hilbert space.
A solution to this problem that enables the use of advanced GAN training methods for training AmbientGANs is described next.

\section{Establishing SOMs by use of Advanced AmbientGANs}
\label{sec:ProAmGAN}
\begin{figure}[H]
\centering
 \includegraphics[width=0.7\linewidth]{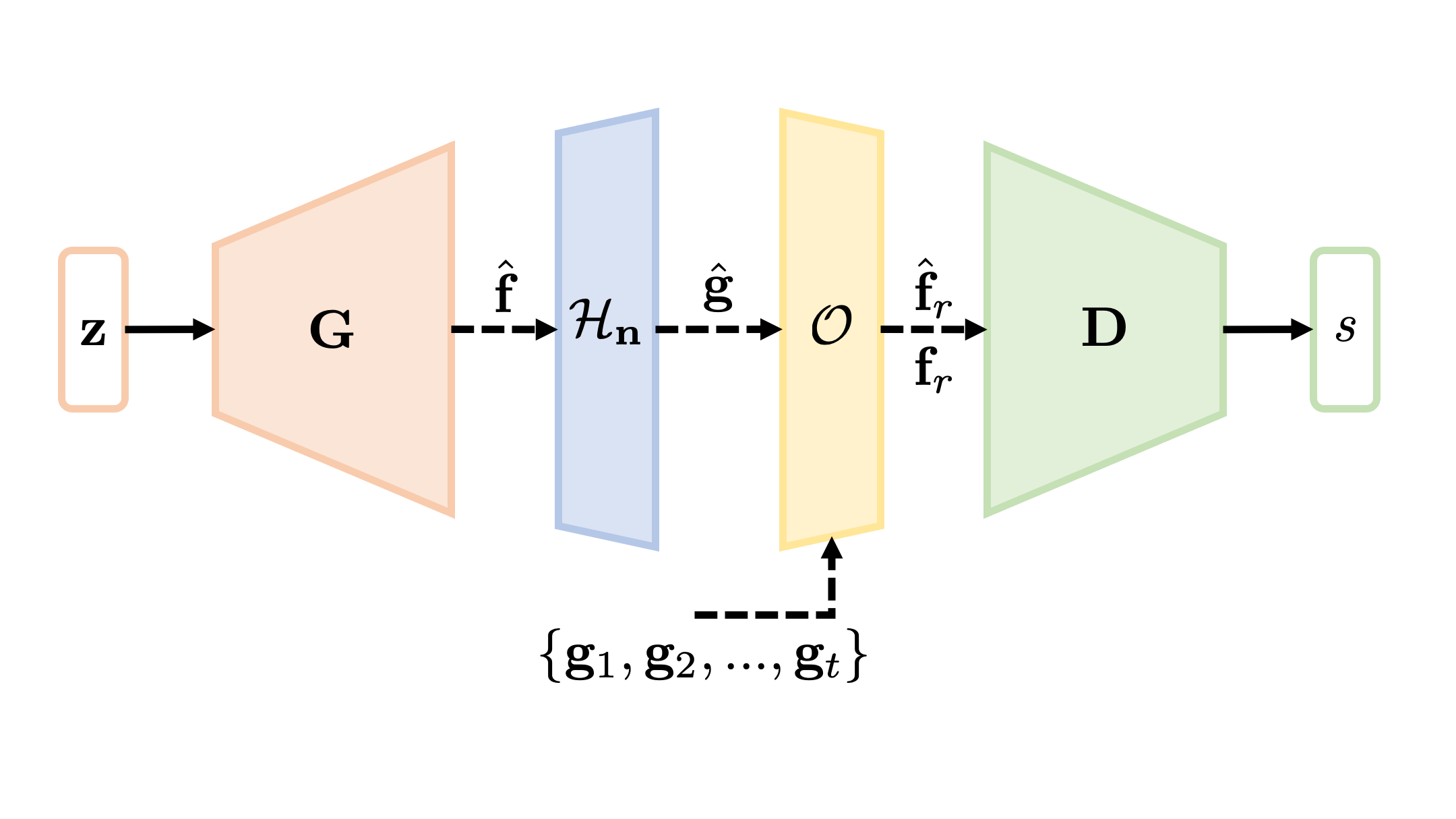}
   \vspace{-0.8cm}
  \caption{An illustration of the proposed modified AmbientGAN architecture. Any advanced GAN architecture employing a progressive growing training procedure can be employed in this framework.}
  \label{fig:arc_PAGAN}
  \end{figure}

In order to train the AmbientGAN 
by use of advanced GAN training methods that employ the progressive growing approach, such as ProGAN and Style-based GANs,
an image reconstruction operator
 $\mathcal{O}$: $\mathbb{R}^{M} \rightarrow \mathbb{R}^N$ 
 is included in the AmbientGAN architecture.
 The discriminator is trained to distinguish between the real reconstructed images $\vec{f}_r = \mathcal{O}(\vec{g})$
and the fake reconstructed images $\hat{\vec{f}}_r = \mathcal{O}(\hat{\vec{g}})$.
In this way, 
the generator and the discriminator are associated with images in the same Hilbert space,
which enables the use of advanced GAN training methods to train AmbientGANs.
This advanced AmbientGAN training strategy is illustrated in Fig.~\ref{fig:arc_PAGAN}.
 \vspace{-0.5cm}
  Given a training dataset that comprises measured data $\vec{g}$,
 a set of reconstructed images $\vec{f}_{r}$
 is computed by applying
 the reconstruction operator $\mathcal{O}$ to the measured data $\vec{g}$.
The generator is trained with the discriminator through an adversarial process to generate objects $\hat{\vec{f}} = \mathbf{G}(\mathbf{z}; \mathbf{\Theta}_{G})$  that result in (fake) reconstructed images $\hat{\vec{f}}_{r}$ that are
indistinguishable, in distribution, from the real reconstructed images ${\vec{f}}_{r}$.
This adversarial training process can be represented by a two-player minimax game:
\begin{equation} \label{eq:AAGAN}
\min_{\mathbf{\Theta}_G} \max_{\mathbf{\Theta}_D} V(D,G) =  {E_{\vec{f}_r\sim  p({\vec{f}}_r) }} [l\left(D(\vec{f}_r; \mathbf{\Theta}_D)\right)]
 + {E_{\hat{\vec{f}}_r \sim p(\hat{\vec{f}}_r) }} [l(1- D( \hat{\vec{f}}_r; \mathbf{\Theta}_D ) )],
\end{equation}
where $\hat{\vec{f}}_r = \mathcal{O}(\mathcal{H}_\vec{n}(\vec{G}(\vec{z}; \mathbf{\Theta}_G)))$.
As with the original AmbientGAN, when the distribution of objects $p(\vec{f})$ uniquely induces the distribution of reconstructed objects $p(\vec{f}_{r})$, and the generator and the discriminator achieve the global optimum, 
the trained generator can be employed to produce object samples drawn from the distribution $p(\vec{f})$.
 
   \begin{figure}[ht!]
	\centering
	\includegraphics[width=0.92\linewidth]{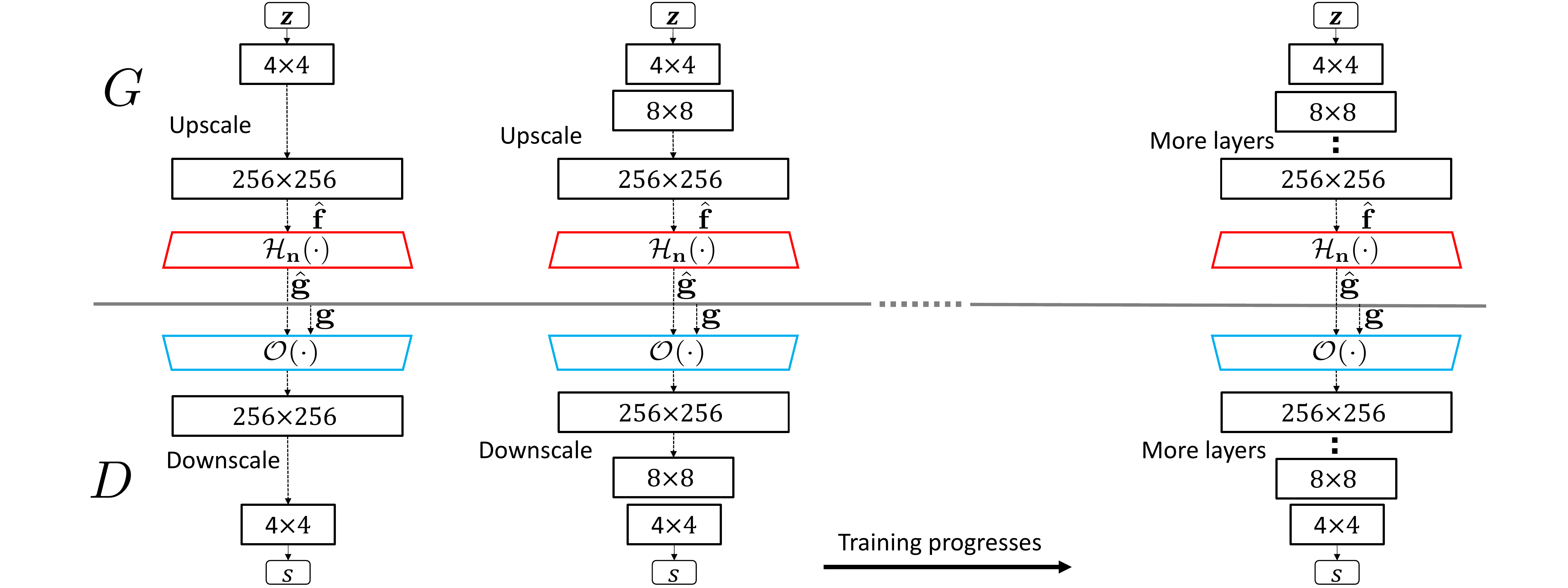}
	\vspace{0.1cm}
	\caption{ProAmGAN training procedure. Initially, the generator and discriminator are trained with low-resolution images. Additional layers in the generator and discriminator are trained by use of higher-resolution versions of the original images when the training advances. More details about the progressive growing method can be found in the original ProGAN paper \cite{karras2018progressive}.}
	\label{fig:proamgan}
\end{figure}
 It should be noted that when the generator and the discriminator are established progressively by use of ProGAN or StyleGAN methods, 
 the generator is initially trained to produce low resolution images that are subsequently upscaled to the original image dimension.
 The measurement operator $\mathcal{H}_\vec{n}$ is subsequently applied to the upscaled images to simulate the measurement data
 and the reconstructed images are produced by use of the reconstruction operator $\mathcal{O}$.
 The reconstructed images are down-sampled and
the discriminator is subsequently trained on the down-sampled (low resolution) reconstructed images.
The generator and the discriminator are progressively trained until the original high-resolution images are achieved.
The training procedure of AmbientGAN that employs progressive growing strategy is illustrated in Fig. \ref{fig:proamgan}.

   \begin{figure}[ht!]
	\centering
	\includegraphics[width=0.92\linewidth]{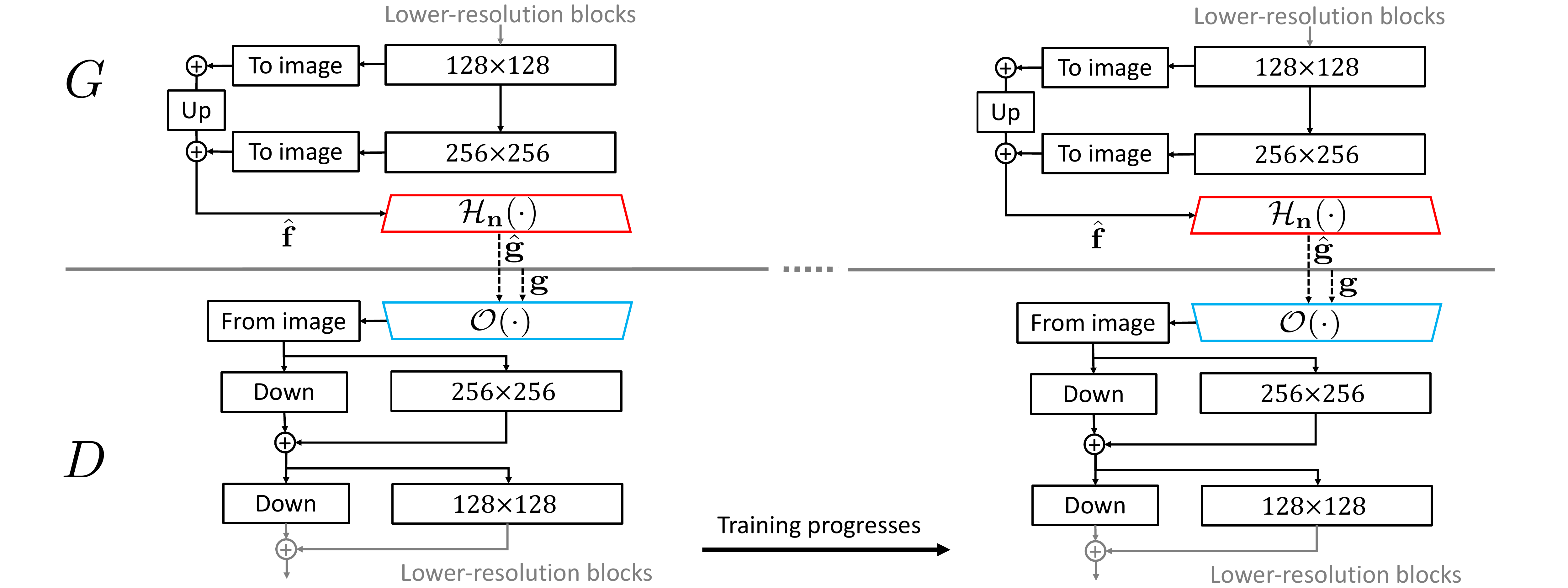}
	\vspace{0.1cm}
	\caption{Training procedure of AmbientGAN that employs the StyleGAN2 architecture. The generator employs skip connections and forms the images by explicitly summing images at different resolutions. The discriminator employs residual connections that can be helpful for performing image classification tasks. 
	{The ``To image'' block corresponds to the convolutional operator that maps hidden feature maps having the spatial dimension of $n\times n$ (e.g., $128\times 128$) to the grayscale image having the dimension of $n\times n$. Similarly, the ``From image'' block corresponds to the convolutional operator that maps the grayscale image to the hidden feature maps. The ``Down'' and ``Up'' blocks denote bilinear down- and up-sampling, respectively.}
	The generator and discriminator are trained without progressive growing {(i.e., the complete architectures of the generator and discriminator are trained during the whole training process.)} More details about the StyeGAN2 architecture can be found in the original StyGAN2 paper \cite{karras2019analyzing}.}
	\label{fig:sty2amgan}
\end{figure}
While the progressive growing strategy has achieved many successes in stabilizing the GAN training for synthesizing high-resolution images,
it can cause certain artifacts in the generated images \cite{karras2019analyzing}.
As mentioned above, the StyleGAN2 that trains a redesigned generator without progressive growing was developed to further improve the synthesized image quality \cite{karras2019analyzing}.  
The new generator employs skip connections to form images that are summation of images with different resolutions. This enables the multi-resolution training of the generator without the explicit use of progressive growing strategy. 
The training of AmbientGANs can be potentially improved further by employing the StyleGAN2 generator and discriminator in the proposed AmbientGAN training framework that is illustrated in Fig. \ref{fig:arc_PAGAN}.
The training procedure of AmbientGAN that employs the StyleGAN2 architecture is illustrated in Fig. \ref{fig:sty2amgan}.

Below, the advanced AmbientGAN that employs the ProGAN was referred to as ProAmGAN and the one that employs the StyleGAN2 was referred to as {Sty2AmGAN}. 
\section{Numerical and experimental studies}
\label{sec:num}

Computer-simulation and experimental studies were conducted to demonstrate the ability of the proposed advanced AmbientGAN methods to establish  SOMs from imaging measurements.
Details regarding the design of these studies are provided below.

\subsection{Stylized imager that acquires fully-sampled data}
\label{sec:MR-full-k-space}
\begin{figure}[H]
   \centering
 \includegraphics[width=0.9\linewidth]{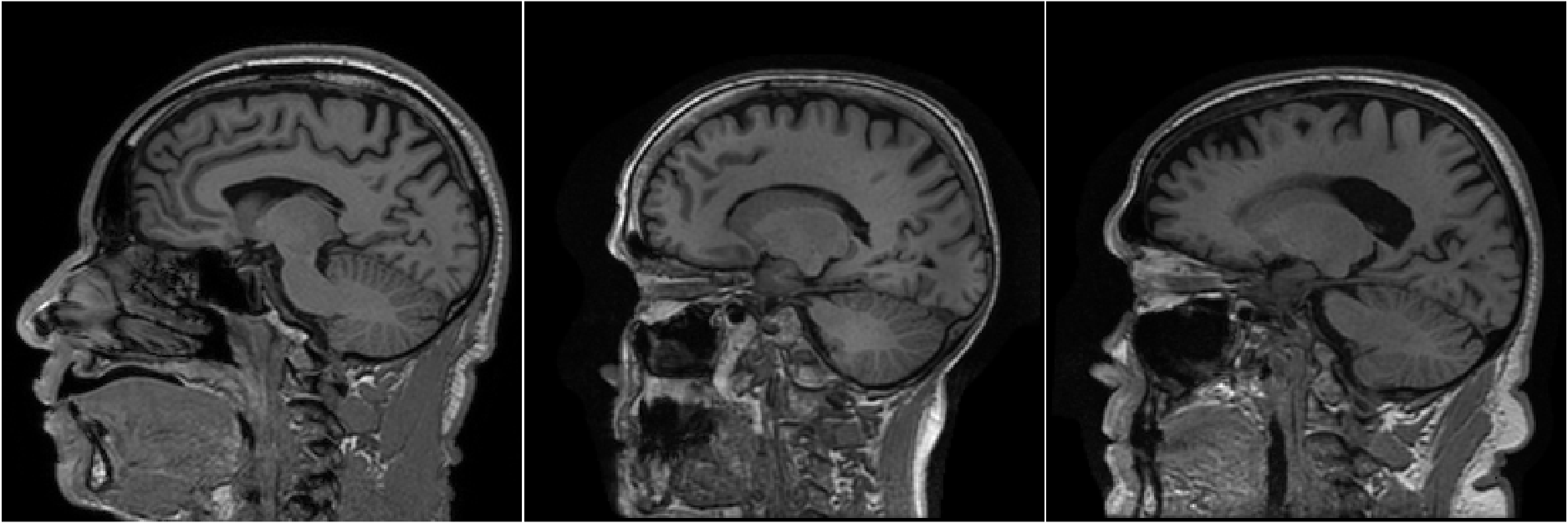}
\vspace{0.2cm}
 \caption{Examples of ground-truth objects $\vec{f}$.}
 \label{fig:real}
\end{figure}

A stylized imaging system that acquires fully-sampled 2D Fourier space (a.k.a., k-space) data was investigated first. 
This imaging system can be described as: 
\begin{equation}\label{eq:mri}
\vec{g} = \mathcal{F}(\vec{f}) + \vec{n}, 
\end{equation}
where $\mathcal{F}$ denotes a 2D discrete Fourier transform (DFT), $\vec{f}$ denotes the discretized object to-be-imaged, and $\vec{n}$
denotes the measurement noise.   While Eq.\ (\ref{eq:mri}) can be interpreted as a simplified model of MRI, it should be noted that here we do not attempt to model the real-world complexities of data-acquisition in MRI.  A situation
where modeling error is present is addressed later in Sec.\ \ref{sec:emulated}.
A collection of clinical brain MR images from the Alzheimer's Disease Neuroimaging Initiative (ADNI) database (\url{adni.loni.usc.edu}) \cite{mueller2005alzheimer} were employed to serve as ground truth objects $\vec{f}$. 
Fifteen thousand sagittal brain slices of dimension $256\times 256$ were selected from this dataset and were normalized to the range between 0 and 1. 
These images were employed to form the collection of ground-truth objects $\vec{f}$. 
Examples of $\vec{f}$ are shown in Fig. \ref{fig:real}.

From the ensemble of objects $\vec{f}$, k-space measurement data were simulated according to Eq. (\ref{eq:mri}).
The measurement noise $\vec{n}$ was modeled by i.i.d. zero mean complex Gaussian distribution with a standard deviation of $\sigma_\vec{n}(\vec{g})$
for both the real and imaginary components.
Different measurement noise levels corresponding to  standard deviations $\sigma_\vec{n}(\vec{g})=4$ and 16 were considered.

From each ensemble of simulated k-space data, 
reconstructed images  $\vec{f}_{r}$ were produced by
acting a 2D inverse discrete Fourier transform (IDFT) operator $\mathcal{F}^{-1}$  to the measured image data $\vec{g}$ and taking the real component: $\vec{f}_{r} = Re(\mathcal{F}^{-1}(\vec{g}))$.
For each noise level,
 {the proposed AmbientGANs were} trained to establish a SOM that characterizes the distribution of objects $\vec{f}$ by use of the ensemble of reconstructed noisy images $\vec{f}_{r}$. 
For comparison, standard (i.e., non-ambient) ProGANs were trained directly by use of reconstructed images $\vec{f}_{r}$. 
In this case,  because the reconstructed images are affected by measurement noise,
the resulting generator will learn to sample from the distribution of noisy reconstructed images instead of the
distribution of (noiseless) objects $\mathbf{f}$.

The Fr\'{e}chet Inception Distance (FID)~\cite{heusel2017gans} score, 
a widely employed metric for assessing generative models, 
was computed to evaluate the performance of the original ProGAN and the proposed AmbientGANs. 
The FID score quantifies the distance between the features extracted by the Inception-v3 network~\cite{szegedy2016rethinking} from the ground-truth (``real") and generated (``fake") objects. 
Lower FID score indicates better quality and diversity of the generated objects. 
The FID scores were computed by use of 15,000 ground-truth objects, 15,000 ProGAN-generated objects, 15,000 ProAmGAN-generated objects {and 15,000 Sty2AmGAN-generated objects}.

As another form of comparison between the ProGAN- and AmbientGANs-generated images, 
the standard deviation of the noise in the generated images $\sigma_{n}(\hat{\vec{f}})$ was estimated.
Specifically, a previously described method \cite{immerkaer1996fast} was applied to 15,000 ProGAN-generated images, 15,000 ProAmGAN-generated images, {and 15,000 Sty2AmGAN-generated images}. 
The average of the estimated standard deviation of the noise in
the ProGAN- and AmbientGANs-generated images were compared.

\subsection{Stylized imager that acquires incomplete data} 
\label{sec:MR_sampling}

Imaging systems sometimes acquire under-sampled, or incomplete, measurement data
to accelerate the data-acquisition process or for other purposes.
In such cases, the imaging operator $\mathbf{H}$ has a non-trivial null space 
and only the measurement component $\vec{f}_{meas} = \mathbf{H}^{\dagger}\mathbf{H}\vec{f}$
can be observed by the imaging system,
where $\mathbf{H}^{\dagger}$ denotes the Moore-Penrose pseudo-inverse of $\mathbf{H}$.
Because of this, it is expected that the performance of an AmbientGAN trained by use of incomplete measurements will be
adversely affected by this information loss. This topic is
investigated below and the
extent to which  ProAmGANs {and Sty2AmGANs} can learn to sample from the distribution of measurement components of an object is demonstrated.

The ensemble of 15,000 clinical MR images that was described in Sec.\ \ref{sec:MR-full-k-space} was employed to serve as ground truth objects.  
Three accelerated data-acquisition designs that under-sample k-space by use of the Cartesian sampling pattern with an acceleration factor (also known as the reduction factor) $R$ of $1.25$, $2$ and $4$ were considered.
The acceleration factor $R$ is defined as the ratio of the amount of fully-sampled k-space data to the amount of k-space data collected in the accelerated data-acquisition process.
For each considered design, a collection of 15,000 measured data $\vec{g}$ were simulated by computing and sampling the k-space data and adding i.i.d. zero mean Gaussian noise with a standard deviation of 4 to both the real and imaginary components.

A stylized imager was considered in which  under-sampled k-space data were acquired and $\mathbf{H}^{\dagger}$  could therefore be computed by applying 
a 2D IDFT to the zero-filled k-space data.
For each data-acquisition design,
reconstructed objects $\vec{f}_{r}$ were produced by acting  $\mathbf{H}^{\dagger}$ on the given measured image data $\vec{g}$.
A ProAmGAN {and a Sty2AmGAN were} subsequently trained to establish a SOM for each data-acquisition design.
 In the training process, $\mathbf{H}$ and $\mathbf{H}^{\dagger}$ were applied to the generator-produced objects as discussed in Sec.\ \ref{sec:ProAmGAN}.
 The FID scores were computed by use of 15,000 ground-truth objects $\vec{f}$ and 15,000 AmbientGANs-generated objects $\hat{\vec{f}}$
 for each data-acquisition design. 
To assess the ability of ProAmGANs {and Sty2AmGANs} to accurately learn the variation in the measurement components of the objects,
 the FID score was computed by use of the ground-truth measurement components $\vec{f}_{meas} = \mathbf{H}^{\dagger}\mathbf{H}\vec{f}$
 and AmbientGANs-generated measurement components $\hat{\vec{f}}_{meas}= \mathbf{H}^{\dagger}\mathbf{H}\hat{\vec{f}}$ for each data-acquisition design.

\subsection{Experimental emulated single-coil MRI data}
\label{sec:emulated}
As a step towards transcending the stylized studies, an emulated set of single-coil knee MRI k-space measurements were also employed to train a ProAmGAN and Sty2AmGAN.  These measurements
were obtained from the NYU fastMRI Initiative database \cite{zbontar2018fastmri}.
The central $256 \times 256$ regions of the k-space were extracted and a total of 11,400 k-space acquisitions were employed for model training.
The reconstructed images were formed as the magnitude of the IDFT of the k-space data.
The magnitude MR images are commonly employed in MRI because they can avoid the phase artifacts that are commonly presented in experimental MR measurement data \cite{gudbjartsson1995rician}.

When training the ProAmGAN and Sty2AmGAN, the canonical measurement model
was assumed: ${\hat{\vec{g}}} =\mathcal{H}_{\mathbf{n}}(\hat{\mathbf{f}})= \mathcal{F}({\hat{\vec{f}}}) + \vec{n}$.
However, when dealing with experimental measurements, the noise model that characterizes $\vec{n}$ is unknown and needs
to be estimated. This was accomplished as follows.
The noise in the (emulated) experimental k-space measurements was assumed to be described
by i.i.d. complex-valued Gaussian random variables; accordingly, the noise in the reconstructed magnitude MR image was  modeled by a Rayleigh distribution \cite{gudbjartsson1995rician}. 
The standard deviation of the measurement noise was subsequently estimated by fitting a Rayleigh distribution to a set of patches, residing outside the support of the object, in the magnitude images that were reconstructed from the noisy k-space measurements. 
The estimated standard deviation specified the k-space noise model in the measurement model above.

In order to train the ProAmGAN and Sty2AmGAN by employing the magnitude MR images as the input to the discriminator, 
care must be taken when computing the simulated reconstructed image $\hat{\vec{f}}_{r}$.
Specifically, 
if the modulus operator, which is denoted as $abs(\cdot)$, is directly applied to the IDFT of the simulated k-space measurements $\hat{\vec{g}}$, i.e., $\hat{\vec{f}}_{r} = abs(\mathcal{F}^{-1}(\hat{\vec{g}})) \equiv abs(\hat{\vec{f}} + \mathcal{F}^{-1}(\vec{n}))$,
the fake magnitude images $\hat{\vec{f}}_{r}$ can be indistinguishable from the real magnitude images $\vec{f}_r$
despite the fact that the corresponding fake objects $\hat{\vec{f}}$ can be negative.
This can prevent the generator from being properly trained for use as a SOM.

To address this issue, 
we computed the fake reconstructed image $\hat{\vec{f}}_{r}$ as:
\begin{equation}
\hat{\vec{f}}_{r} = \hat{\vec{f}} + \vec{e},
\end{equation}
where $\vec{e} = abs(\mathcal{F}^{-1}(\mathcal{F}(ReLU(\hat{\vec{f}})) + \vec{n})) - ReLU(\hat{\vec{f}})$.
Here, $ReLU(\cdot)$ is the component-wise Rectified Linear Unit (ReLU) operator
that outputs the input value if the input value is positive; while if the input value is negative, it outputs 0.
The quantity $\hat{\vec{f}}_{r}$ can be subsequently expressed as:
\begin{equation}
    \hat{\vec{f}}_{r}= 
\begin{cases}
    abs(\mathcal{F}^{-1}(\mathcal{F}(\hat{\vec{f}})+\vec{n}) ),& \text{if } \hat{\vec{f}}\geq 0\\
    \hat{\vec{f}} +abs(\mathcal{F}^{-1}(\vec{n})),              & \text{if } \hat{\vec{f}}< 0
\end{cases}
\end{equation}
In this way, 
fake reconstructed images $\hat{\vec{f}}_r$ that are produced by positive objects
can represent the corresponding magnitude images
while those that are produced by negative objects
cannot represent magnitude images.
Therefore, 
when the training is completed such that the fake reconstructed images $\hat{\vec{f}}_r$ are indistinguishable from the ground-truth magnitude MR images ${\vec{f}}_r$,
the generator would produce non-negative objects.

\subsection{Task-based image quality assessment} 
\begin{figure}[H]
	\centering
	\includegraphics[width=0.9\linewidth]{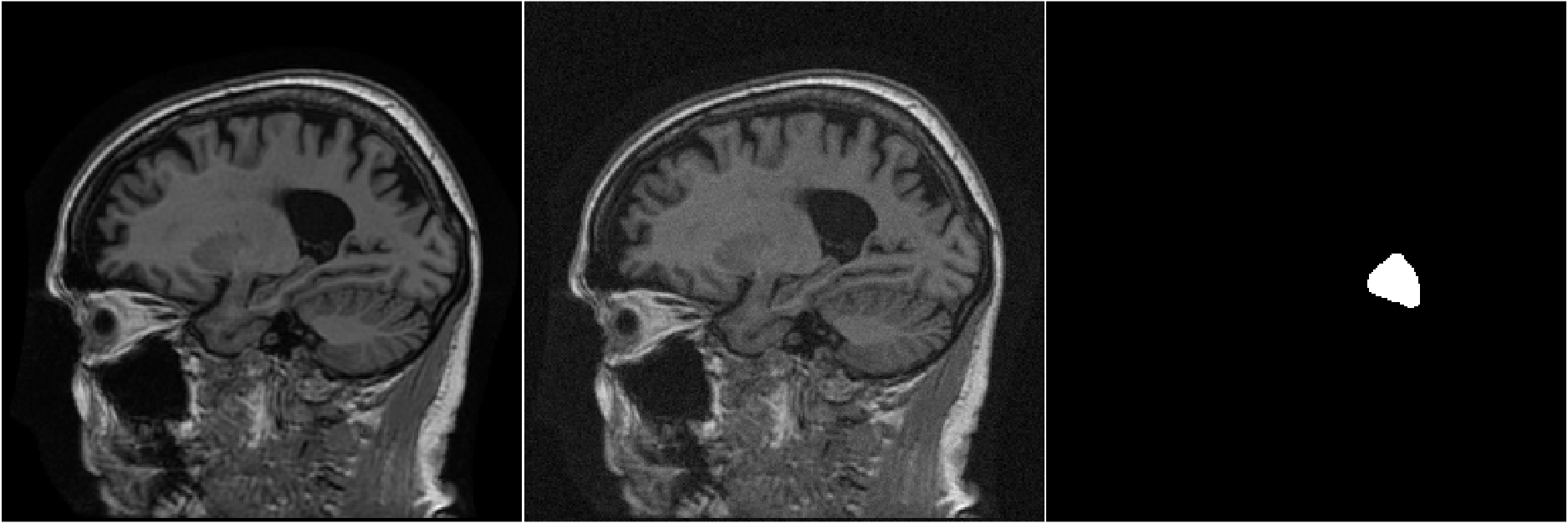}
	\vspace{0.2cm}
	\caption{From left to right: an example of a ground truth, or ``real", object $\vec{f}$, the corresponding noisy signal-absent image $\vec{g}_t$, and the considered signal $\vec{s}$.}
 \label{fig:signal}
\vspace{-0.2cm}
\end{figure}

The generative models established by use of the ProGANs, ProAmGANs, {and Sty2AmGANs} in the stylized numerical studies described in Sec. \ref{sec:MR-full-k-space} and  Sec. \ref{sec:MR_sampling}  were further evaluated 
by use of objective measures of IQ.
To accomplish this, 
a signal-known-exactly and background-known-statistically (SKE/BKS) binary classification task was considered. 
A Hotelling observer was employed to classify  noisy images $\vec{g}_t$ as satisfying either a signal-absent hypothesis ($H_0$) or signal-present hypothesis~($H_1$): 
\begin{subequations}
\label{eq:imgH_s}
\begin{align}
H_{0}:&\ \mathbf{g}_t = \mathbf{f} + \mathbf{n}_t, \\
H_{1}:&\  \mathbf{g}_t = \mathbf{f} + \mathbf{s} + \mathbf{n}_t,
\end{align}
\end{subequations}
where $\mathbf{s}$ denotes {the considered signal placed at the fixed location} and $\vec{n}_t$ is i.i.d. zero-mean Gaussian noise having the standard deviation of $2\%$.
Two studies were conducted in which the background objects $\mathbf{f}$  corresponded to ground truth brain images or synthetic images produced by use of an AmbientGAN.
As such, this study sought to determine whether the GAN-generated objects could `fool' the Hotelling observer on the specified detection task.
An example of the ``real" object $\vec{f}$, the corresponding noisy signal-absent image $\vec{g}_t$, and the considered signal are shown in Fig.~\ref{fig:signal}.

\vspace{-0.3cm}

The considered signal detection task was performed on a region of interest (ROI) of dimension of $64\times 64$ pixels centered at the signal location.
The signal-to-noise ratio of the Hotelling observer (HO) test statistic $\text{SNR}_{HO}$ was employed as 
the figure-of-merit for assessing the image quality\cite{barrett2013foundations}:
\begin{equation}
\label{eq:snr}
  \text{SNR}_{HO} = \sqrt{{\vec{s}_{ROI}}^T\mathbf{K}^{-1}{\vec{s}_{ROI}}}, 
\end{equation}
where $\vec{s}_{ROI} \in \mathbb{R}^{4096\times 1}$ denotes the vectorized signal image in the ROI, 
and $\mathbf{K} \in \mathbb{R}^{4096\times 4096}$ denotes the covariance matrix corresponding to the ROIs in the noisy MR images.  
When computing 
$\text{SNR}_{HO}$, $\mathbf{K}^{-1}$ was calculated by use of a covariance matrix decomposition~\cite{barrett2013foundations}.
The values of $\text{SNR}_{HO}$ were computed by use of 15,000 generated objects produced by each trained generative model.
They were compared to the $\text{SNR}_{HO}$ computed by use of 15,000 ground truth objects.

\subsection{Training details}
All ProGAN, ProAmGAN, and Sty2AmGAN models were trained by use of Tensorflow\cite{abadi2016tensorflow} on 2 NVIDIA Quadro RTX 8000 GPUs.
The Adam algorithm~\cite{kingma2014adam}, which is a stochastic gradient algorithm, was employed as the optimizer in the training process.
To implement the ProAmGAN, the ProGAN code ( \url{https://github.com/tkarras/progressive_growing_of_gans}) was modified according to Fig.~\ref{fig:arc_PAGAN}.
Specifically, for each considered imaging system, the corresponding measurement operator 
was applied to the generator-produced images for simulating the measurement data and
the reconstruction operator was applied to the measurement data for producing the reconstructed images used as the input to the discriminator.
The default ProGAN architecture with the latent space having the dimensionality of 512 and the initial image resolution of $4\times 4$
was employed to implement the ProAmGANs for the considered numerical studies.
Additional details about the ProGAN architecture and the progressive growing training method can be found in the literature \cite{karras2018progressive}.

The Sty2AmGAN was implemented by modifying the StyleGAN2 code (\url{https://github.com/NVlabs/stylegan2})
by augmenting the StyleGAN2 with the measurement operator $\mathcal{H}_n$
 and the reconstruction operator $\mathcal{O}$ according to Fig. \ref{fig:arc_PAGAN}.
 For the considered experimental study,
 the default StyleGAN2 architecture (i.e., ``config F"  \cite{karras2019analyzing}) with the input latent space having the dimensionality of 512
 was employed to implement the Sty2AmGAN.
 Additional details regarding the StyleGAN2 architecture and the corresponding training strategy can be found in the literature \cite{karras2019analyzing}.

{During the training of ProAmGANs and Sty2AmGANs, the latent vectors were sampled from the standard normal distribution to generate fake images. 
We visually examined these generated fake images and stopped the training after these images possessed a plausible visual quality or did not visually improve significantly.
We acknowledge that this stopping rule is subjective, and it remains an open problem to quantitatively evaluate AmbientGANs to be applied in situations where ground-truth objects are not accessible.
}

\section{Results}
\label{sec:result}
\subsection{Stylized imager that acquires fully-sampled data}
\begin{figure}[ht!]
     \centering
        \includegraphics[width=0.9\linewidth]{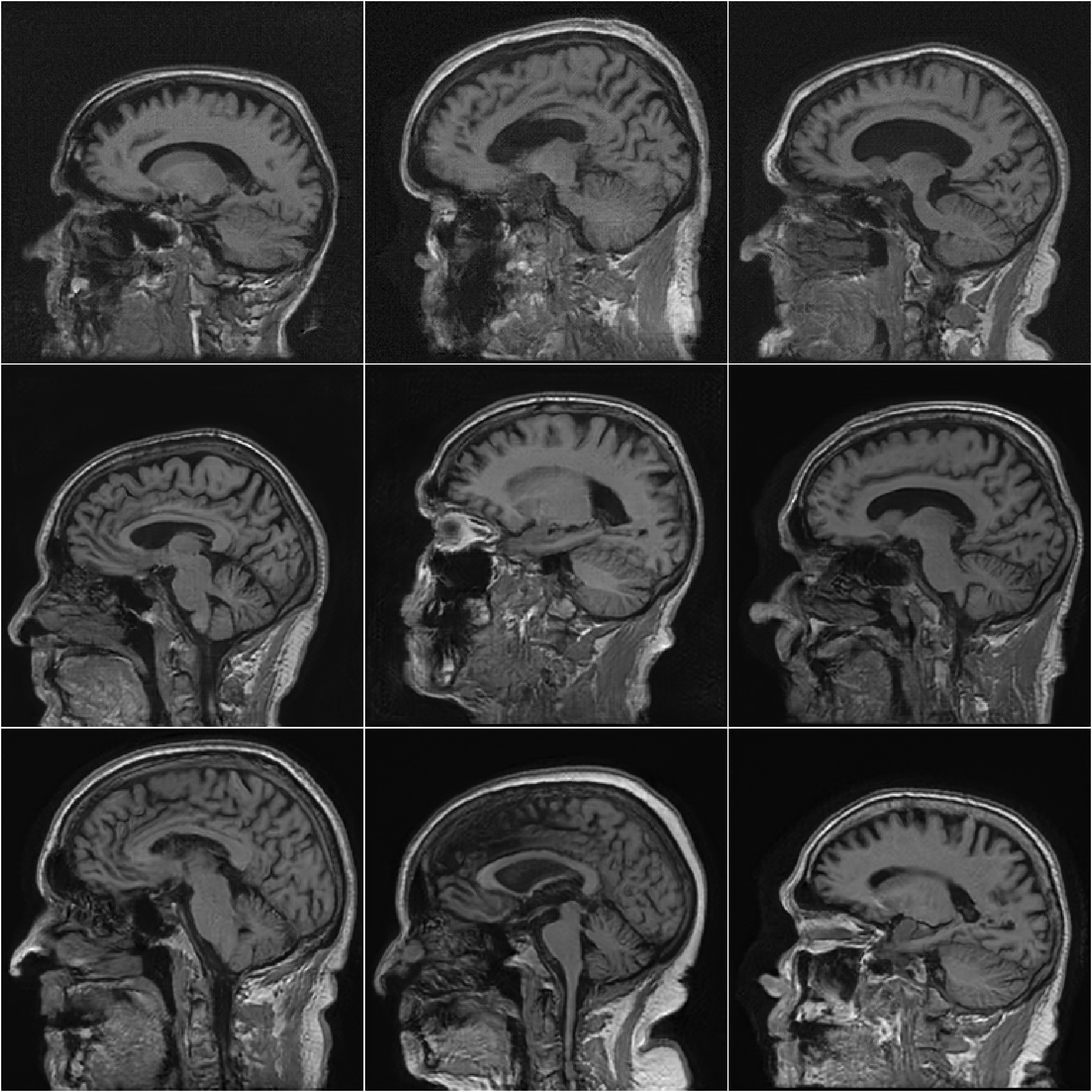}
       \vspace{0.2cm}
        \caption{{ProGAN-generated (top row), ProAmGAN-generated (middle row), and Sty2AmGAN-generated (bottom row) images corresponding to $\sigma_{n}(\vec{g})=4$. }}
         \label{fig:4n_fakes}
        \end{figure}

\begin{figure}[ht!]
\centering
        \includegraphics[width=0.9\linewidth]{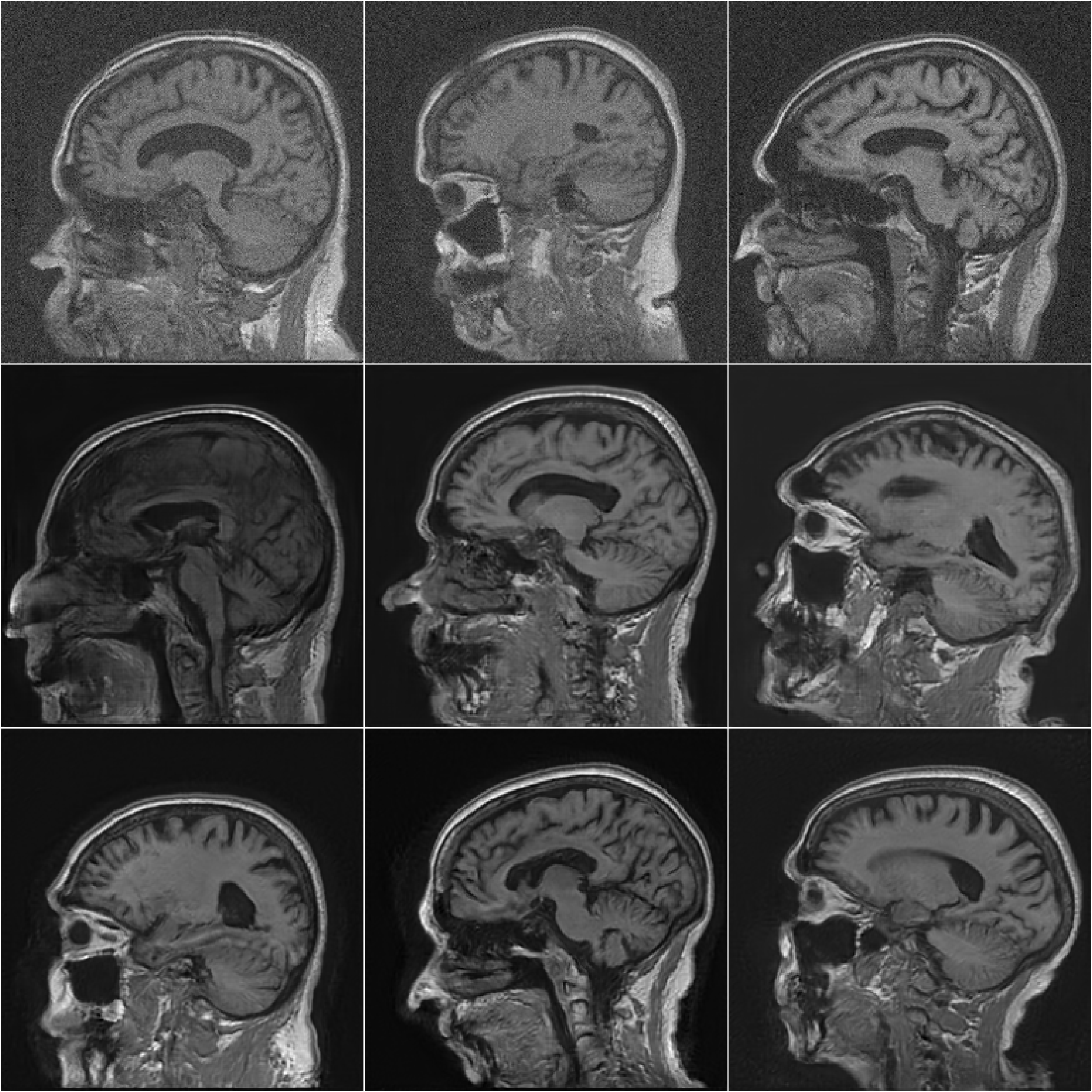}
        \vspace{0.2cm}
  \caption{{ProGAN-generated (top row), ProAmGAN-generated (middle row), and Sty2AmGAN-generated (bottom row) images corresponding to $\sigma_{n}(\vec{g})= 16$. }}
  \label{fig:16n_fakes}
\end{figure}

{Images that were synthesized by use of the advanced-AmbientGANs and ProGANs that were trained by use of fully-sampled noisy k-space data or images reconstructed from them, respectively, are shown in Figs. {\ref{fig:4n_fakes}} and {\ref{fig:16n_fakes}}.  These correspond to measurement noise levels of $4$ and $16$, respectively.}
\vspace{-0.1cm}
        
It is observed that the ProGAN-generated images contain significant noise when $\sigma_n(\vec{g})=16$, while the
ProAmGAN {and Sty2AmGAN} generated clean images that do not contain significant noise. This demonstrates the ability of the ProAmGAN {and Sty2AmGAN} to mitigate measurement noise when establishing SOMs.

\begin{figure}[ht!]
\centering
        \includegraphics[width=0.9\linewidth]{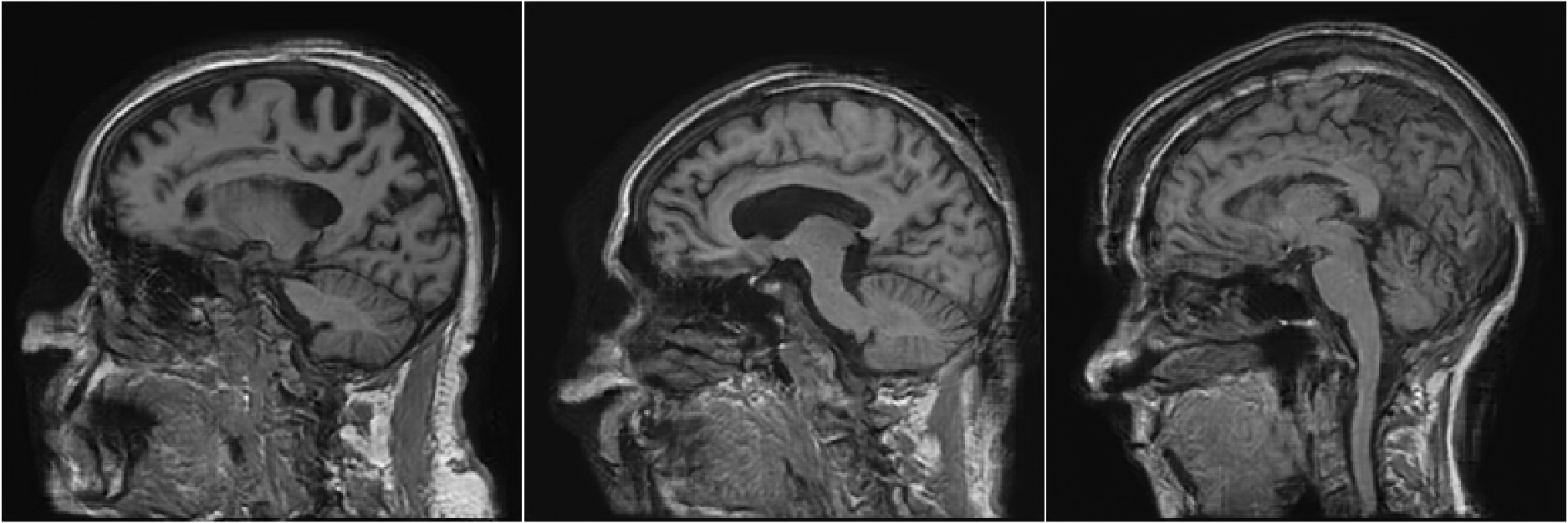}
        \vspace{0.2cm}
  \caption{{ProAmGAN-produced images having significant artifacts near the skulls that were not observed in Sty2AmGAN-produced images. These images were produced by the ProAmGAN corresponding to $\sigma_{n}(\vec{g})=4$.}}
  \label{fig:ProAmGAN_artifact}
\end{figure}

The FID scores, estimated standard deviation of the noise in the generated images $\sigma_n(\hat{\vec{f}})$, and $\text{SNR}_{HO}$ were evaluated for { the ProGANs, ProAmGANs, and Sty2AmGANs}.
These metrics are shown in Table. \ref{table:1}.
The ProAmGANs produced FID scores that were smaller than those produced by the ProGANs, which indicates that the ProAmGANs outperformed the ProGANs.
{Additionally, Sty2AmGANs can further improve the synthesized image quality and produced FID scores smaller than the ProAmGANs.} 
{It was also observed that ProAmGANs can produce images having  artifacts that did not appear in Sty2AmGAN-produced images. Examples of such images are shown in Fig. {\ref{fig:ProAmGAN_artifact}}.}
The estimated standard deviation of the noise in the ProGAN-generated images increased nearly linearly as the standard deviation of measurement noise was increased; while
the estimated standard deviation of the noise in the {ProAmGAN and Sty2AmGAN-generated images} were almost unchanged.
The $\text{SNR}_{HO}$ values corresponding to the ProGANs had negative biases to the reference value  that were computed by use of ground-truth objects, and this negative bias became more significant as the measurement noise level increased.
This is because the ProGANs capture both the object variability and the noise randomness, instead of object variability alone, which degrades the estimated observer performance. 
The $\text{SNR}_{HO}$ values corresponding to the ProAmGANs {and Sty2AmGANs} were closer to the reference value.
\vspace{-0.05cm}
\renewcommand{\arraystretch}{1.2}
\begin{table}[H]
\begin{adjustbox}{width=1\columnwidth,center}
\begin{tabular}{l|c|c||c|c||c|c}
\hline
\hline
                                                                       & \multicolumn{2}{c||}{\textbf{ProGAN}} & \multicolumn{2}{c||}{\textbf{ProAmGAN}} & \multicolumn{2}{c}{\textbf{Sty2AmGAN}} \\ \cline{2-7} 
                                                                       & $\sigma_{n}(\vec{g}) = 4$   & $\sigma_{n}(\vec{g}) = 16$          & $\sigma_{n}(\vec{g}) = 4$    & $\sigma_{n}(\vec{g}) = 16$          & $\sigma_{n}(\vec{g}) = 4$    & $\sigma_{n}(\vec{g}) = 16$  \\ \hline
\begin{tabular}[c]{@{}l@{}}FID ($\hat{\vec{f}}$) \   \end{tabular}                          & 25.31  & 146.92  & 17.60  & 37.66   & 13.84  & 21.24 \\ \hline
\begin{tabular}[c]{@{}l@{}} $\sigma_n(\hat{\vec{f}})$  \end{tabular}       & 1.77\%   & 6.43\%   & 0.59\%   & 0.51\%     & 0.59\%   & 0.59\%\\ \hline
\begin{tabular}[c]{@{}l@{}}$\text{SNR}_{HO}$ \end{tabular} & 1.62 & 1.10 & 1.73 & 1.78      & 1.70 & 1.66 \\ \hline
\end{tabular}
\end{adjustbox}
\vspace{0.2cm}
\caption{The FID score of the objects, the estimated noise standard deviation, and the $\text{SNR}_{HO}$ (the reference value 1.72) 
		 corresponding to the objects produced by the {ProGANs, ProAmGANs, and Sty2AmGANs} that were trained with fully-sampled noisy k-space measurement data.}
\label{table:1}
\end{table}
\renewcommand{\arraystretch}{1}

\begin{figure}[H]
   \centering
 \includegraphics[width=0.9\linewidth]{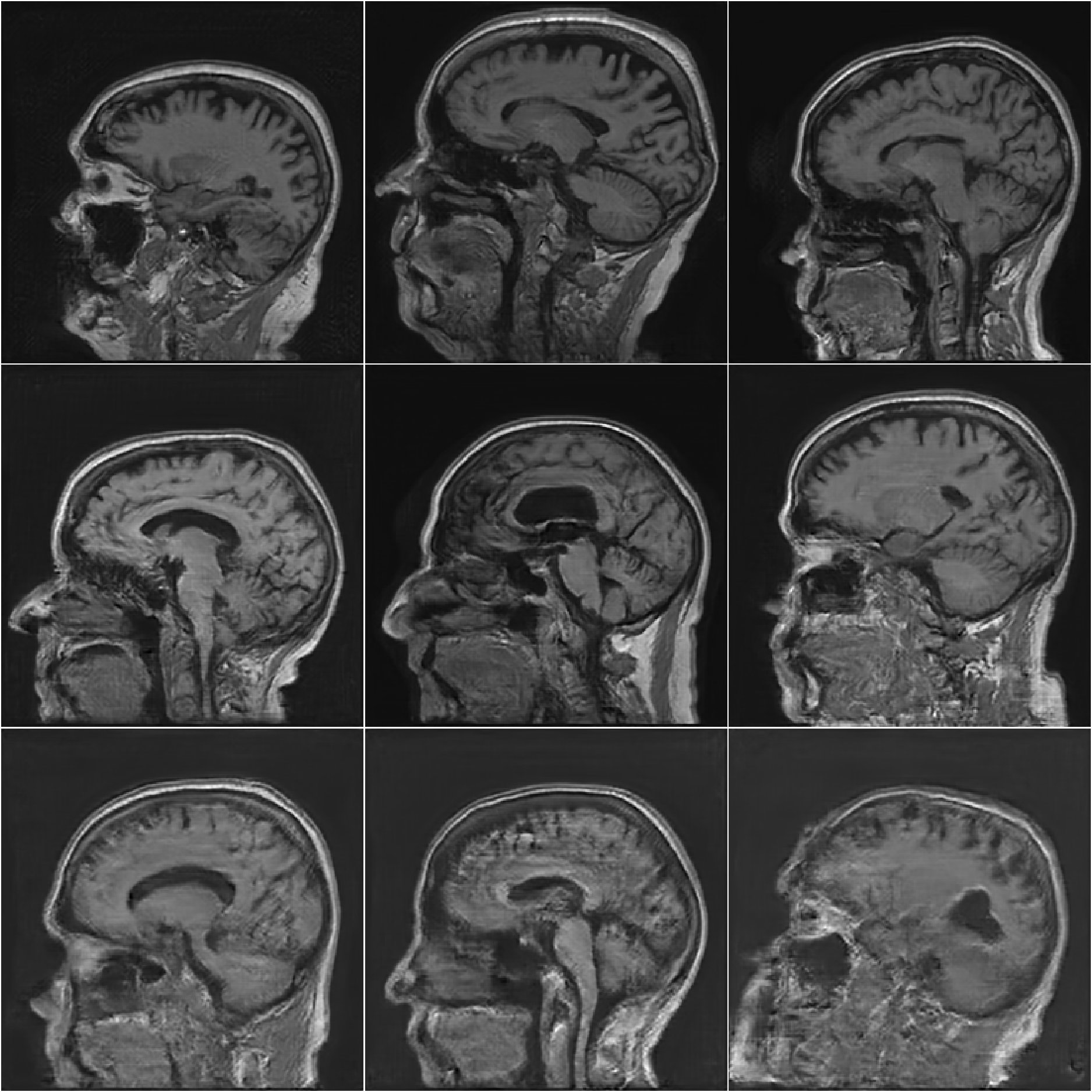}
 \vspace{0.2cm}
 \caption{Top row: ProAmGAN-generated images corresponding to the k-space sampling acceleration factor $R$ of 1.25. Middle row: The corresponding images for $R=2$. Bottom row: The corresponding images for $R=4$.}
 \label{fig:null_fakes}
\end{figure}

\subsection{Stylized imager that acquires incomplete data}

\begin{figure}[ht!]
   \centering
 \includegraphics[width=0.9\linewidth]{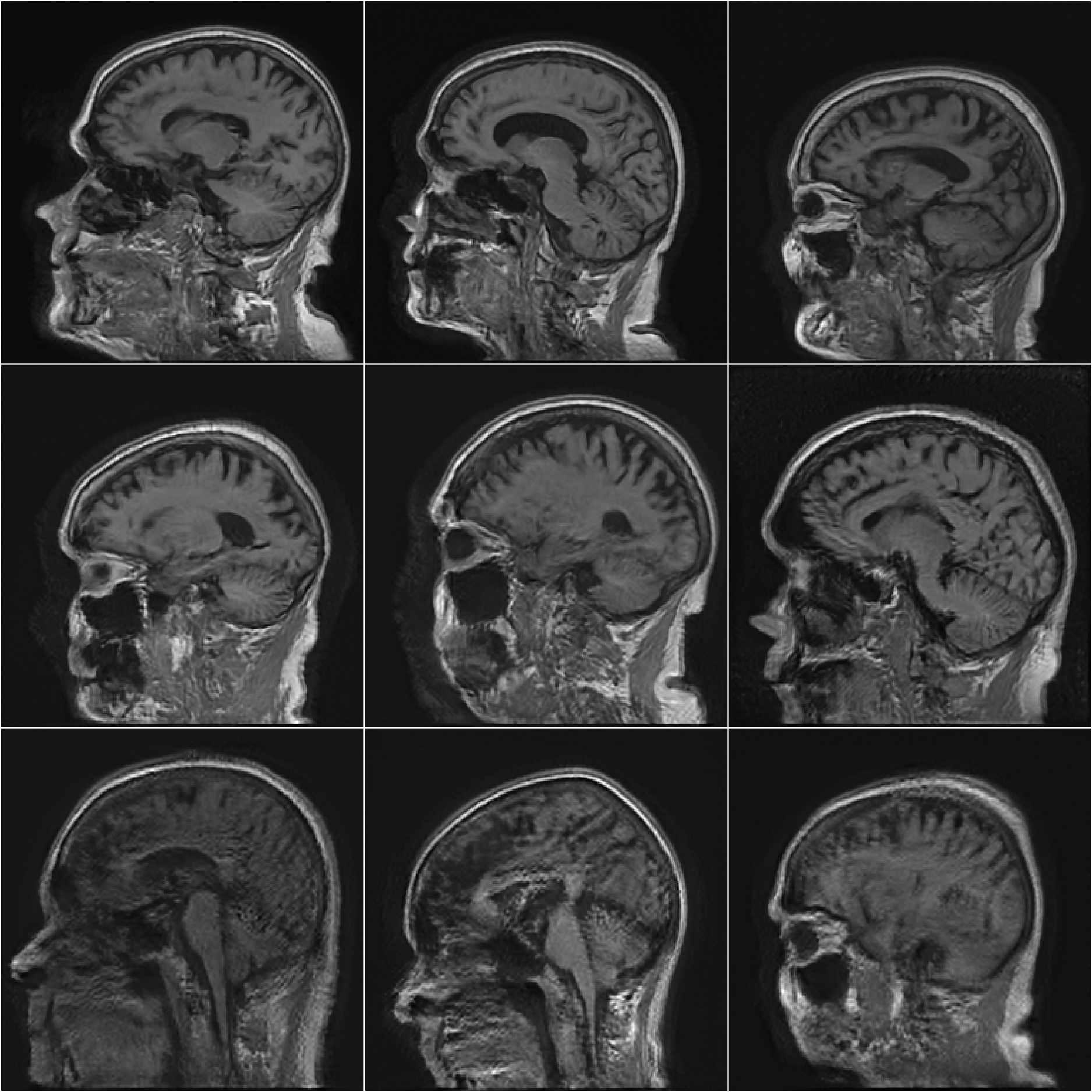}
 \vspace{0.2cm}
 \caption{{Top row: Sty2AmGAN-generated images corresponding to the k-space sampling acceleration factor $R$ of 1.25. Middle row: The corresponding images for $R=2$. Bottom row: The corresponding images for $R=4$.}}
 \label{fig:null_fakes_sty}
\end{figure}

Images  that  were  synthesized  by  use  of  the  ProAmGANs {and Sty2AmGANs} that were trained by use of under-sampled k-space measurement data acquired with different acceleration factors are  shown in Fig. \ref{fig:null_fakes}
{and Fig. {\ref{fig:null_fakes_sty}}, respectively. }
The images produced by the ProAmGANs {and Sty2AmGANs} corresponding to the acceleration factor $R=1.25$ and 2 are visually plausible;
while
when the acceleration factor was increased to $4$, 
the generated-images were obviously contaminated by artifacts and some structures were distorted.
This demonstrates that the ProAmGAN {and Sty2AmGAN} were adversely affected by the incompleteness of the measurement data acquired by imaging systems having non-trivial null-space.

The quantitative metrics that include FID scores and $\text{SNR}_{HO}$ are summarized in Table. \ref{table:2}.
{The FID scores produced by the Sty2AmGANs were smaller than those produced by the ProAmGANs. This indicates that the Sty2AmGANs outperformed the ProAmGANs in terms of FID scores.}
{For both the ProAmGANs and Sty2AmGANs, }
the FID scores corresponding to the generated objects $\hat{\vec{f}}$
were increased when the acceleration factor $R$ increased, which
indicates that the ProAmGAN {and Sty2AmGAN were} detrimentally affected by the null space of the imaging operator.
However,
the FID scores corresponding to the measurement components of the generated objects $\hat{\vec{f}}_{meas}$ were not significantly affected,
which suggests that the ProAmGAN {and Sty2AmGAN} can reliably learn the distribution of the measurement components of the objects.
{The $\text{SNR}_{HO}$ values produced by the ProAmGANs and Sty2AmGANs increased when the k-space sampling acceleration factor $R$ increased.
This suggests that the ability of AmbientGANs to learn object variation that limits observer performance can be decreased when the null space of the imaging operator becomes large.}

\renewcommand{\arraystretch}{1.2}
\begin{table}[ht!]
\begin{adjustbox}{width=0.8\columnwidth,center}
\begin{tabular}{l|c|c|c||c|c|c}
\hline
\hline
							& \multicolumn{3}{c||} {\textbf{ProAmGAN}} & \multicolumn{3}{c}{\textbf{Sty2AmGAN}} \\ \cline{2-7}                                                             
                                                                       & $R = 1.25$ 	& $R = 2$   & $R = 4$           & $R = 1.25$ 	& $R = 2$   & $R = 4$\\ \hline
\begin{tabular}[c]{@{}l@{}}FID ($\hat{\vec{f}}$) \   \end{tabular}              & 20.64 & 39.25 & 118.27   & 16.40 & 37.76 & 109.41\\ \hline
\begin{tabular}[c]{@{}l@{}} FID ($\hat{\vec{f}}_{meas}$) \end{tabular}       & 12.83 & 13.25 & 8.96   & 10.52  & 12.51 &11.80\\ \hline
\begin{tabular}[c]{@{}l@{}}$\text{SNR}_{HO}$ \end{tabular} & 1.75 & 1.80 & 1.84   & 1.66  & 1.73 & 1.77\\ \hline
\end{tabular}
\end{adjustbox}
 \vspace{0.2cm}
\caption{The FID score of the objects, the FID score of the measurement components, and the $\text{SNR}_{HO}$ (reference value 1.72) corresponding to the objects produced by the ProAmGANs {and Sty2AmGANs} that were trained with under-sampled k-space data with different acceleration factors.}
\label{table:2}
\end{table}
\renewcommand{\arraystretch}{1}

\subsection{Experimental emulated single-coil MRI data}
\begin{figure}[H]
   \centering
 \includegraphics[width=0.9\linewidth]{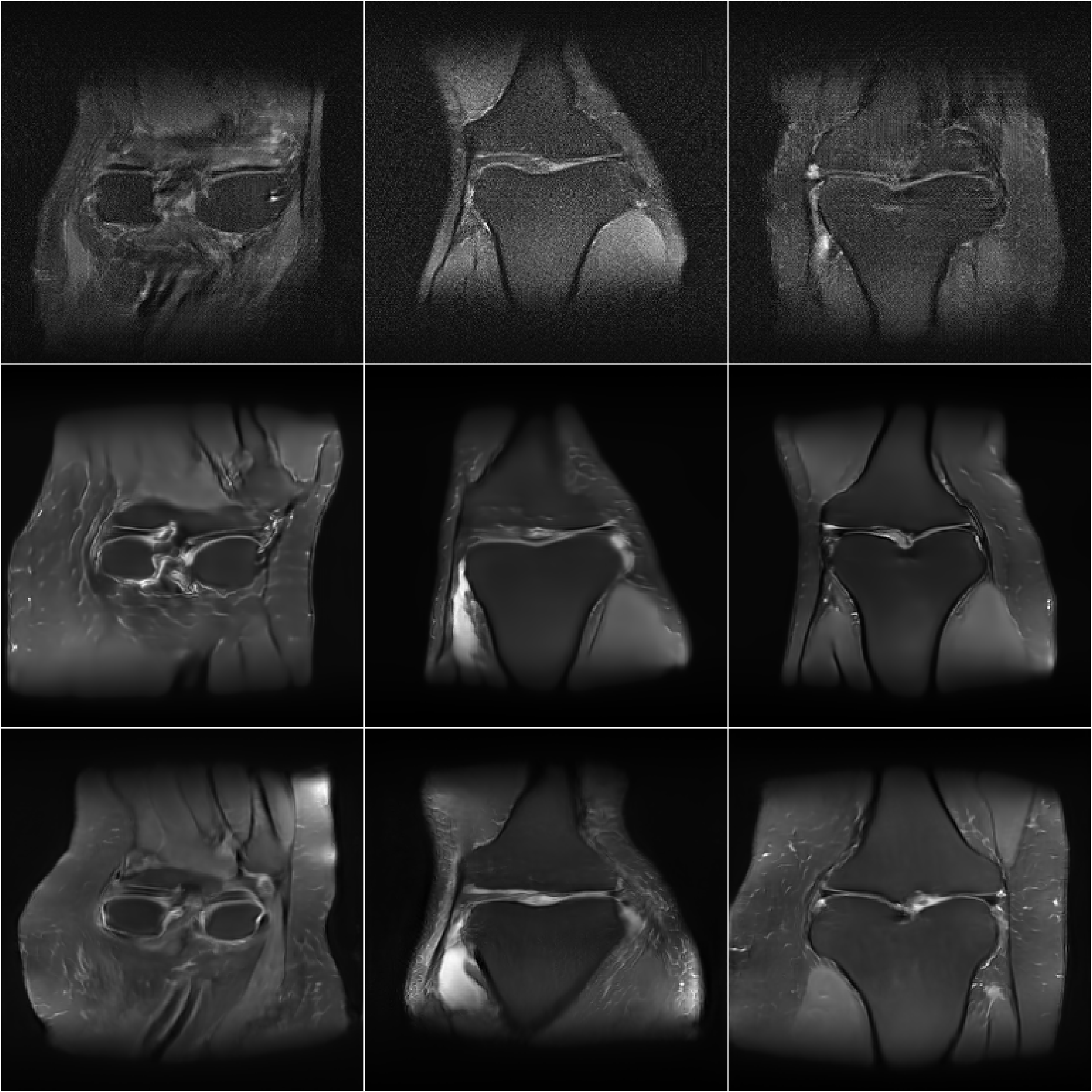}
 \vspace{0.1cm}
 \caption{Results from emulated single-coil MRI data. Top row: ProGAN-generated images. Middle row: ProAmGAN-generated images. Bottom row: Sty2AmGAN-generated images.}
 \label{fig:NYU_fakes}
\end{figure}

\begin{figure}[H]
   \centering
 \includegraphics[width=0.9\linewidth]{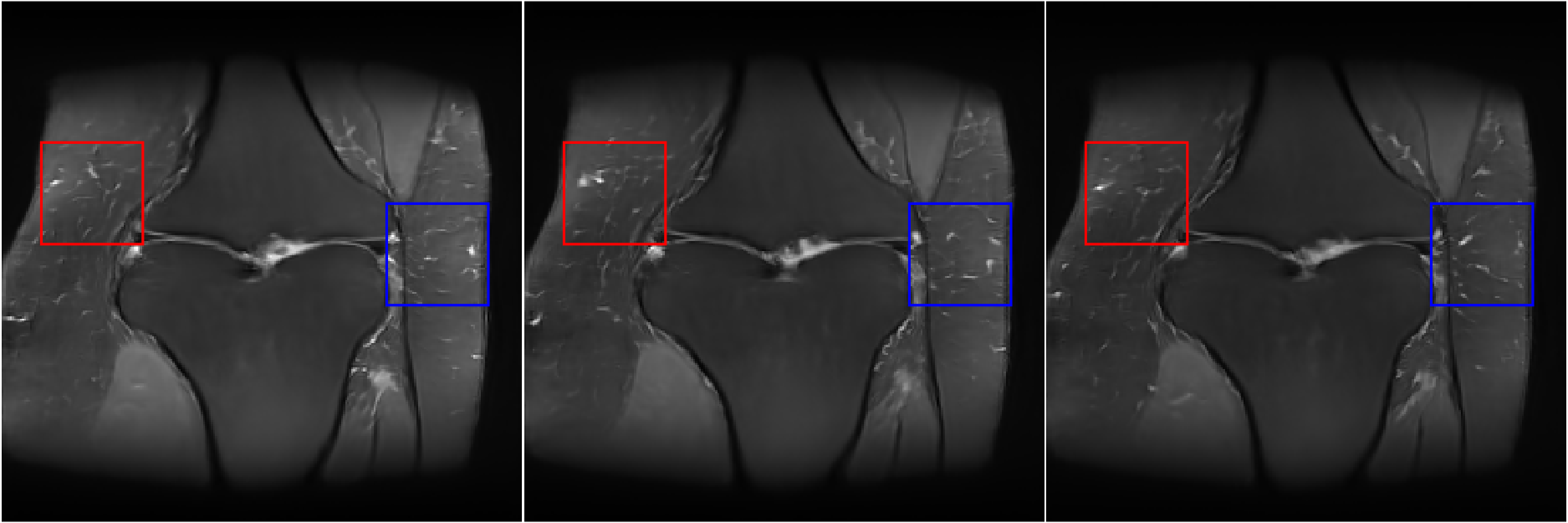}
 \vspace{0.1cm}
 \caption{{Sty2AmGAN-generated objects having  similar large-scale structures but different fine-scale subcutaneous fat textures. Textures in the red and blue rectangles have different appearances.}}
 \label{fig:NYU_fakes_scales}
\end{figure}

Images produced by the ProGAN, ProAmGAN, and Sty2AmGAN are shown in the top row, middle row, and bottom row of Fig. \ref{fig:NYU_fakes}, respectively.
The ProGAN-produced images were contaminated by noise because the ProGAN was trained directly by use of noisy reconstructed images.
Both the ProAmGAN and Sty2AmGAN produced images that did not appear to be degraded by noise, which demonstrates the ability of advanced AmbientGAN strategies to mitigate the measurement noise when establishing an SOM.
The Sty2AmGAN can further improve the training of the AmbientGAN for establishing the SOM.
For example, the images produced by the ProAmGAN were  more blurred than those
produced by the 
 Sty2AmGAN.
 Because the ground-truth objects corresponding to the synthesized images were not accessible in this experimental study, only a subjective visual assessment was performed.
The style-based generator used in Sty2AmGAN can provide additional ability to control scale-specific image features \cite{zhou2021advancing,kelkar2021prior}.
{To demonstrate this, as shown in Fig. {\ref{fig:NYU_fakes_scales}}, we controlled the style-based generator of the Sty2AmGAN to produce knee images having similar large-scale  structures but different fine-scale subcutaneous fat textures. These images were produced by use of 
the same latent vector 
but different latent noise maps that are extra inputs to the style-based generator. More details about the latent noise maps and the scale-specific manipulation of the style-based generators can be found in the references} \cite{karras2019style, karras2019analyzing}.


\section{Discussion and conclusion}
\label{sec:conclusion}
It is known that it is important to address object variability when computing objective measures of image quality for use in imaging system characterization or optimization.
When computer-simulation studies are employed, SOMs are the means by which this can be accomplished. However, establishing realistic SOMs from experimental image data has remained challenging and few methods are available to accomplish this.

Motivated by this need,  a methodology for training AmbientGANs by use of medical image data was proposed in this study. The trained generator of the AmbientGAN represents the sought-after SOM.
The proposed methodology
enables the use of advanced GAN training methods and architectures in the AmbientGAN training
and therefore permits the AmbientGAN to be applied to realistically sized medical image data.
To demonstrate this, two specific advanced AmbientGANs were considered: ProAmGANs and Sty2AmGANs.

Stylized numerical studies were systematically conducted in which Sty2AmGANs and ProAmGANs were trained on simulated measurement data corresponding to an object ensemble and a variety of stylized imaging systems.
Both visual examinations and quantitative analyses, including task-specific validations, demonstrated that the proposed ProAmGANs and Sty2AmGANs
hold promise to establish realistic SOMs from imaging measurements.
Additionally, an experimental study was conducted in which the ProAmGAN and Sty2AmGAN were trained on emulated experimental MRI measurement data. This demonstrated the effectiveness of the methods under less stylized conditions in which modeling error was present.

In addition to objectively assessing imaging systems and data-acquisition designs,
the SOMs established by the proposed advanced AmbientGAN methods can be employed to regularize image reconstruction problems. 
Recent methods have been developed for regularizing image reconstruction problems based on GANs such as Compressed Sensing using Generative Models (CSGM)\cite{bora2017compressed} and image-adaptive GAN-based reconstruction methods (IAGAN)\cite{hussein2019image, bhadra2020medical}. 
Sty2AmGANs can also be used for prior image-constrained reconstruction\cite{kelkar2021prior}.
{Furthermore, the
established SOMs may be employed} to produce clean reference images for training deep neural networks for 
solving other image-processing problems such as image denoising\cite{zhang2017beyond, li2021assessing} and image super-resolution\cite{dong2014learning}.
{However, it should be noted that the AmbientGANs-established SOMs were not uninfluenced by the imaging systems.
As shown in the numerical studies, the image quality of the AmbientGAN-generated images was affected to varying extents when different levels of measurement noise and different degrees of incompleteness of the measurement data were considered.
Additionally, the generated images can be contaminated by artifacts and distorted structures. 
The artifacts generated by AmbientGANs and the impact of imaging systems on the training of AmbientGANs may limit the use of the proposed AmbientGANs in certain medical imaging applications.
It will be important to investigate the extent to which the AmbientGANs can be successfully applied for solving specific medical imaging problems.}

There remain additional topics for future investigation. 
We have conducted a preliminary objective assessment of the AmbientGANs
by use of the Hotelling Observer \cite{barrett2013foundations, zhou2019learning_HO} and a binary signal detection task.
{In this preliminary study,  the generated images were objectively assessed by use of   binary signal detection studies.}
It will be important to validate the SOMs established by use of the proposed methods  when  clinically relevant tasks are addressed by a variety of numerical observers such as the Ideal Observer {that deploys higher-order statistical information of images}\cite{zhou2018learning, zhou2019approximating, zhou2020markov,zhou2020approximating} and anthropomorphic observers {that mimic human performance}\cite{massanes2017evaluation}.
{Moreover, it will also be important to validate the AmbientGANs-established SOMs by use of other image features such as variations of certain textures and shape distributions of different organ structures of the synthesized objects.}
To implement the proposed AmbientGAN methods, the imaging forward operator needs to be accurately modeled.
{Additionally, in practice, medical imaging measurement data are frequently acquired under inhomogeneous imaging conditions.}
It will be important to investigate the impact of the modeling error of the imaging forward operator {and the inhomogeneous imaging conditions} on the ability of AmbientGANs to establish SOMs for certain observers and tasks.

\section{Acknowledgements}
This research was supported in part by NIH grants EB028652, NS102213, EB023045 and EB020604, Cancer Center at Illinois seed grant, and DoD Award E01 W81XWH-21-1-0062.
Data collection and sharing for this project was funded by the Alzheimer's Disease Neuroimaging Initiative (ADNI) (National Institutes of Health Grant U01 AG024904) and DOD ADNI (Department of Defense award number W81XWH-12-2-0012).


\bibliography{AmbientGAN}   
\bibliographystyle{spiejour}   


\vspace{2ex}\noindent\textbf{Dr.\ Weimin Zhou} is a Postdoctoral Scholar in the Department of Psychological \& Brain Sciences at University of California, Santa Barbara (UCSB). Before joining UCSB in 2020, he was a research assistant in the Department of Biomedical Engineering at Washington University in St.~Louis~(WashU) and a visiting scholar in the Department of Bioengineering at University of Illinois at Urbana-Champaign (UIUC). Dr.~Zhou earned his Ph.D. in Electrical Engineering at WashU in 2020. His research focuses on biomedical imaging, machine learning, image science, and vision science. Dr. Zhou is the recipient of the SPIE Community Champion and the SPIE Medical Imaging Cum Laude Poster Award. 

\vspace{1ex}
\noindent\textbf{Sayantan Bhadra} is a Ph.D. candidate in the Department of Computer Science and Engineering, Washington University in St. Louis, and a visiting research scholar in the Department of Bioengineering, University of Illinois at Urbana-Champaign. He received the B.E. degree in Electronics and Telecommunication Engineering from Jadavpur University in 2016. His research interests include computational imaging, inverse problems and machine learning. 

\vspace{1ex}
\noindent\textbf{Dr.\ Frank J. Brooks} received his PhD in Physics from Washington University in Saint Louis (WU). His dissertation specialization was in biophysics, rate equations and statistical physics. He applied these skills to various problems in statistical image analysis, image acquisition and radio-tracer kinetics at the Department of Radiation Oncology at the Washington University School of Medicine (WUSM) and again at the Department of Radiology at the Mallinckrodt Institute of Radiology, also at WUSM. In 2017, he accepted a position at the WU Department of Biomedical Engineering where he began research at the intersection of machine learning and imaging science. In 2019, Dr.\ Brooks began his current research professorship at the Grainger College of Engineering Department of Bioengineering at the University of Illinois Urbana-Champaign where he is a senior member of the Computational Imaging Science Laboratory. His current research interests are in the creation of stochastic object models and the evaluation of generative adversarial networks.

 \vspace{1ex}
\noindent\textbf{Dr.\ Hua Li} is Professor in the Department of Radiation Oncology at the Washington University in St. Louis.
She is a member of Cancer Center at Illinois. Her research work focuses on four topics: deep learning for clinical decision-making, medical imaging
and image restoration for clinical applications, task-based medical imaging and image restoration
assessment and optimization, and improvement of radiation treatment on cancer patients. Her
research work is supported by NCI and industry funds. She serves as one of the Deputy Editors for
the journal of Medical Physics, reviewer for a set of journals and NIH study sections. She is the
lead inventor of several US patents.

\vspace{1ex}
\noindent\textbf{Dr.\ Mark A. Anastasio} is the Donald Biggar Willett Professor in Engineering and the Head of the Department of Bioengineering at the University of Illinois at Urbana-Champaign (UIUC).  He is a Fellow of the SPIE, American Institute for Medical and Biological Engineering (AIMBE) and International Academy of Medical and Biological Engineering (IAMBE).  Dr.\ Anastasio’s research broadly addresses computational image science, inverse problems in imaging, and machine learning for imaging applications.  He has contributed broadly to emerging biomedical imaging technologies that include photoacoustic computed tomography, ultrasound computed tomography, and X-ray phase-contrast imaging. His work has been supported by numerous NIH grants and he served for two years as the Chair of the NIH EITA Study Section.  


\end{spacing}
\end{document}